\documentclass[vecphys]{svmult}

\usepackage{makeidx}         
\usepackage{graphicx}        
\usepackage{multicol}        
\usepackage[bottom]{footmisc}

\makeindex     

\begin{document}

\title*{GRB afterglows in the ELT era}
\titlerunning{GRB afterglows}
\author{
David Alexander Kann\inst{1} \and
Sylvio Klose\inst{1}
}
\authorrunning{D. A. Kann \& S. Klose}

\institute{Th\"uringer Landessternwarte Tautenburg, Germany
\texttt{kann@tls-tautenburg.de, klose@tls-tautenburg.de}
}

\maketitle


Afterglow phenomenology on a statistical basis is a substantial tool to get insight into the physical processes at work (cf. Panaitescu et al. 2006; Zhang 2006). Since several years we have been undertaken such an approach by gathering and analyzing the largest possible optical/NIR data set in a systematic way (Zeh et al. 2004, 2006; Kann et al. 2006, 2007, 2008). In Zeh et al. (2004) we analyzed all optical afterglows searching for supernova light appearing at late times and discovered that the data indicate that all long bursts are related to supernova explosions. In Zeh et al. (2006) and Kann et al. (2006) we investigated the light curve shape and the spectral energy distribution of all optical afterglows known in the pre-Swift era in a systematic way.
Finally, in the most recent and most comprehensive publications (Kann et al. 2007, 2008) we used our data base to discuss the properties of the afterglows of short bursts in comparison to the long burst sample, as well as comparing the long GRB afterglows of the pre-Swift with those of the Swift era.

The plot shows the light curves of a sample of optical afterglows of long GRBs from pre-Swift (gray) and the Swift era (black) selected for good light curve coverage and known redshift, complete until the end of August 2007. The data are corrected for Galactic extinction and, where possible, for host galaxy contribution, but otherwise as observed. In red, we plot the light curves of short GRBs,
in all cases, afterglows were detected (square data points), but additional upper limits are given too (downward pointing triangles). Clearly, the afterglows of short GRBs are much fainter than those of long GRBs, and high S/N or high resolution spectroscopy of these dim cosmic beacons will require the light-gathering power of an ELT.

The inset figure shows early light curves of GRBs that had rapid spectroscopic observations, within 0.1 days. The fastest observations were obtained with VLT UVES in Rapid Response Mode mode. This mode, which allows the follow-up of transients within just minutes, and which will become available for other instruments at the VLT in the next years, will make this observatory competitive far into the ELT era as a powerful tool to study the early evolution of the afterglows of the most powerful explosions in the universe.


\begin{figure}
\centering
\includegraphics[height=11.5cm]{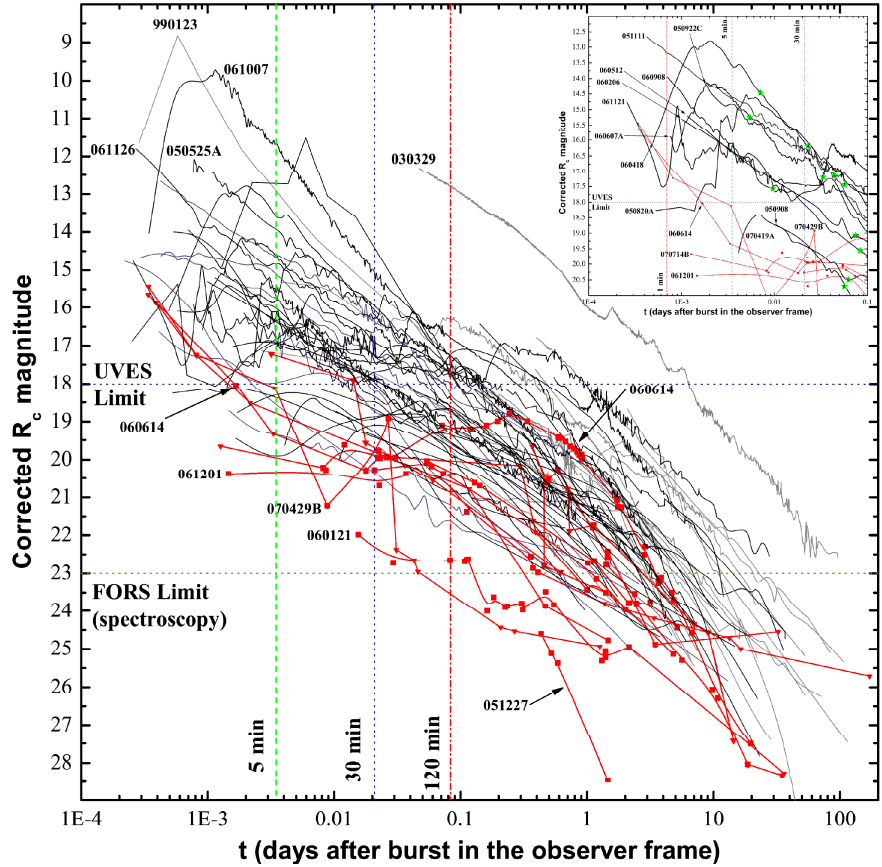}
\caption{GRB optical afterglow light curves with well-sampled data up to the end of August 2007. Pre-Swift long GRB afterglows are grey, Swift era ones are black. Short GRB afterglows are red, with detections marked by squares, and additional upper limits by triangles.The inset shows early light curves of long GRBs with rapid spectroscopic follow-up.}
\end{figure}

\end{document}